\documentclass[english,showpacs,floatfix]{revtex4}
\usepackage{amsfonts}
\usepackage{amssymb,epsf}
\usepackage{latexsym}

\usepackage[dvips]{graphicx}
\usepackage{longtable}
\usepackage{amsmath,amssymb}
\usepackage{dcolumn}
\usepackage{latexsym}
\usepackage{babel}
\usepackage[latin1]{inputenc}
\begin{document}
\title{Rotating Black Holes in Einstein-Maxwell-Dilaton Gravity }
\author{Ahmad Sheykhi \footnote{sheykhi@mail.uk.ac.ir}}
\address{Department of Physics, Shahid Bahonar University, P.O. Box 76175-132, Kerman, Iran\\
         Research Institute for Astronomy and Astrophysics of Maragha (RIAAM), Maragha, Iran}
\begin{abstract}
We present a new class of slowly rotating black hole solutions in
$(n+1)$-dimensional $(n\geq3)$ Einstein-Maxwell-dilaton gravity in
the presence of Liouville-type potential for the dilaton field and
an arbitrary value of the dilaton coupling constant. Because of
the presence of the dilaton field, the asymptotic behaviour of
these solutions are neither flat nor (A)dS. In the absence of a
dilaton field, our solution reduces to the $(n+1)$-dimensional
Kerr-Newman modification thereof for small rotation parameter
\cite{Aliev2}. We also compute the angular momentum and the
gyromagnetic ratio of these rotating dilaton black holes.
\end{abstract}
\maketitle %%%%%%%%%%%%%%%

\section{Introduction}
The motivation for studying higher dimensional black holes
originates from developments in string/M-theory, which is believed
to be the most promising approach to quantum theory of gravity in
higher dimensions. It was argued that black holes may play a
crucial role in the analysis of dynamics in higher dimensions as
well as in the compactification mechanisms. In particular to test
novel predictions of string/M-theory, microscopic black holes may
serve as good theoretical laboratories \cite{Strom,Strom1}.
Another motivation for studying higher dimensional black holes
comes from the braneworld scenarios, as a new fundamental scale of
quantum gravity. For a while it was thought that the extra spatial
dimensions would be of the order of the Planck scale, making a
geometric description unreliable, but it has recently been
realized that there is a way to make the extra dimensions
relatively large and still be unobservable. This is if we live on
a three dimensional surface (brane) in a higher dimensional
spacetime (bulk) \cite{RS,DGP}. In such a scenario, all
gravitational objects  such as black holes are higher dimensional.
One of the most interesting phenomena predicted within this
scenario is the possibility of the formation in accelerators of
higher dimensional black holes smaller than the size of extra
dimensions \cite{Dim}.

The pioneering study on higher dimensional black holes was done by
Tangherlini several decades ago who found the analogues of
Schwarzschild and Reissner-Nordstrom solutions in higher
dimensions \cite{Tan}. These solutions are static with spherical
topology. The rotating black hole solutions in higher dimensional
Einstein gravity was found by Myers and Perry \cite{Myer}. These
solutions (\cite{Myer}) have no electric charge and can be
considered as a generalization of the four dimensional Kerr
solutions. Besides, it was confirmed by recent investigations that
the gravity in higher dimensions exhibits much richer dynamics
than in four dimensions. For example, there exists a black ring
solution in five dimensions with the horizon topology of $S^2
\times S^1$ \cite{Emp} which can carry the same mass and angular
momentum as the Myers-Perry solution, and consequently the
uniqueness theorem fails in five dimensions. While the nonrotating
black hole solution to the higher-dimensional Einstein-Maxwell
gravity was found several decades ago \cite{Tan}, the counterpart
of the Kerr-Newman solution in higher dimensions, that is the
charged generalization of the Myers-Perry solution in
$(n+1)$-dimensional Einstein-Maxwell gravity, still remains to be
found analytically. Indeed, the case of charged rotating black
holes in higher dimensions has been discussed in the framework of
supergravity theories and string theory
\cite{Gau,Yo,Cvetic0,Cvetic1,Cvetic2,Cvetic3}. Recently, charged
rotating black hole solutions in $(n+1)$-dimensional
Einstein-Maxwell theory with a single rotation parameter in the
limit of slow rotation has been constructed in \cite{Aliev2} (see
also \cite{Aliev3,kunz1}). More recently a class of charged slowly
rotating black hole solutions in Gauss-Bonnet gravity has been
presented in \cite{KimCai}.

It is also of great interest to generalize the study to the
dilaton gravity. A dilaton is a kind of scalar field which appears
in the low-energy effective action of string theory and can be
coupled to gravity and gauge fields \cite{Wit1}. When a dilaton is
coupled to Einstein-Maxwell theory, it has profound consequences
for the black hole solutions. This fact may be seen in the case of
four dimensional rotating Einstein-Maxwell-dilaton (EMd) black
holes which does not possess the gyromagnetic ratio $g=2$ of
Kerr-Newman black hole \cite{Gib,Fr,Hor1,Ghosh}. Therefore, the
study on the rotating solutions of Einstein-Maxwell gravity in the
presence of a dilaton field is well motivated. Of particular
interest is to investigate the effect of the dilaton field on the
physical properties of the solutions. The appearance of the
dilaton field changes the asymptotic behavior of the solutions to
be neither asymptotically flat nor (A)dS. A motivation to
investigate non-asymptotically flat, non-asymptotically (A)dS
solutions of Einstein gravity is that these might lead to possible
extensions of AdS/CFT correspondence. Indeed, it has been
speculated that the linear dilaton spacetimes, which arise as
near-horizon limits of dilatonic black holes, might exhibit
holography \cite{Ahar}. Another motivation is that such solutions
may be used to extend the range of validity of methods and tools
originally developed for, and tested in the case of,
asymptotically flat or asymptotically (A)dS black holes. It has
been shown that the counterterm method inspired by the AdS/CFT
correspondence can be applied successfully to the computation of
the conserved quantities of non-asymptotically (A)dS black
holes/branes (see e.g. \cite{DF,SDRP,SDR,DHSR}). While exact
static dilaton black hole/string solutions of EMd gravity have
been constructed in
\cite{CDB1,CDB2,Ast1,MNR,MW,PW,CHM,Cai,Clem,Sheykhi}, exact
rotating black holes solutions with curved horizons have been
obtained only for some limited values of the dilaton coupling
constant \cite{Fr,kun,kunz2,Bri}. For general dilaton coupling
constant, the properties of  charged rotating dilaton black holes
only with infinitesimally small charge \cite{Cas} or small angular
momentum have been investigated \cite{ShR1,Hor1,Ghosh,Shi,ShR2}.
When the horizons are flat, rotating solutions of EMd gravity with
Liouville-type potential in four \cite{DF} and $(n+1)$-dimensions
have also been constructed \cite{SDRP}. These solutions
(\cite{DF,SDRP}) are not black holes and describe charged rotating
dilaton black strings/branes. It is worth noting that these
rotating solutions (\cite{DF,SDRP}) are basically obtained by a
Lorentz boost from corresponding static ones; they are equivalent
to static ones locally, although not equivalent globally. So far,
charged rotating dilaton black hole solutions with curved
horizons, for an arbitrary value of the dilaton coupling constant
in $(n+1)$-dimensional EMd theory have not been constructed. In
this paper, as a new step to shed some light on this issue for
further investigation, we present a class of rotating solutions in
$(n+1)$-dimensional EMd theory. These solutions describe an
electrically charged, slowly rotating dilaton black holes with
curved horizons and an arbitrary value of the dilaton coupling
constantin in $(n+1)$ dimensions. We shall also investigate the
effects of the dilaton field as well as the rotation parameter on
the physical quantities such as temperature, entropy, angular
momentum and the gyromagnetic ratio of these rotating dilaton
black holes.

%%%%%%%%%%%%%%%%%%%%%%%%%%%%%%%%%%%%%%%%%%%%%%%%%%%%%%%%%%%
\section{Basic equations and solutions}

The action of $(n+1)$-dimensional $(n\geq3)$ Einstein-Maxwell
gravity coupled to a dilaton field can be written
\begin{eqnarray}
S &=&\frac{1}{16\pi }\int d^{n+1}x\sqrt{-g}\left(
\mathcal{R}\text{ }-\frac{4}{n-1}(\nabla \Phi )^{2}-V(\Phi
)-e^{-4\alpha \Phi /(n-1)}F_{\mu \nu }F^{\mu \nu }\right)   \nonumber \\
&&-\frac{1}{8\pi }\int_{\partial \mathcal{M}}d^{n}x\sqrt{-\gamma
}\Theta (\gamma ),  \label{act1}
\end{eqnarray}
where ${\cal R}$ is the Ricci scalar curvature, $\Phi$ is the
dilaton field and $V(\Phi)$ is a potential for $\Phi$. $\alpha $
is a constant determining the strength of coupling of the scalar
and electromagnetic field, $F_{\mu \nu }=\partial _{\mu }A_{\nu
}-\partial _{\nu }A_{\mu }$ is the electromagnetic  field tensor
and $A_{\mu }$ is the electromagnetic potential. The last term in
Eq. (\ref{act1}) is the Gibbons-Hawking boundary term which is
chosen such that the variational principle is well-defined. The manifold $%
\mathcal{M}$ has metric $g_{\mu \nu }$ and covariant derivative
$\nabla _{\mu }$. $\Theta $ is the trace of the extrinsic
curvature $\Theta ^{ab}$ of any boundary(ies) $\partial
\mathcal{M}$ of the manifold $\mathcal{M}$, with induced metric(s)
$\gamma _{ab}$. In this paper, we consider the action (\ref{act1})
with a Liouville type potential,
\begin{equation}
V(\Phi )=2\Lambda e^{2\zeta \Phi},  \label{v1}
\end{equation}
where $\Lambda $ and $\zeta$ are arbitrary constants. The
equations of motion can be obtained by varying the action
(\ref{act1}) with respect to the gravitational field $g_{\mu \nu
}$, the dilaton field $\Phi $ and the gauge field $A_{\mu }$ which
yields the following field equations
\begin{equation}
\mathcal{R}_{\mu \nu }=\frac{4}{n-1}\left( \partial _{\mu }\Phi
\partial _{\nu }\Phi +\frac{1}{4}g_{\mu \nu }V(\Phi )\right)
+2e^{-4\alpha \Phi /(n-1)}\left( F_{\mu \eta }F_{\nu }^{\text{
}\eta }-\frac{1}{2(n-1)}g_{\mu \nu }F_{\lambda \eta }F^{\lambda
\eta }\right) ,  \label{FE1}
\end{equation}
\begin{equation}
\nabla ^{2}\Phi =\frac{n-1}{8}\frac{\partial V}{\partial \Phi
}-\frac{\alpha }{2}e^{-{4\alpha \Phi }/({n-1})}F_{\lambda \eta
}F^{\lambda \eta }, \label{FE2}
\end{equation}
\begin{equation}
\nabla _{\mu }\left( e^{-{4\alpha \Phi }/({n-1})}F^{\mu \nu
}\right) =0.  \label{FE3}
\end{equation}
We would like to find $(n+1)$-dimensional rotating solutions of
the above field equations. For infinitesimal rotation, we can
solve Eqs. (\ref{FE1})-(\ref{FE3}) to first order in the angular
momentum parameter $a$. Inspection of the $(n+1)$-dimensional Kerr
solutions shows that the only term in the metric changes to $O(a)$
is $g_{t\phi}$. Similarly, the dilaton field does not change to
$O(a)$ and $A_{\phi}$ is the only component of the vector
potential that changes. Therefore, for infinitesimal angular
momentum up to $O(a)$, we can take the following form of the
metric
\begin{eqnarray}\label{metric}
ds^2 &=&-U(r)dt^2+{dr^2\over U(r)}- 2 a f(r)\sin^{2}{\theta}dt
d{\phi}\nonumber \\
 &&+ r^2 R^2(r)\left(d\theta^2 + \sin^2\theta d\phi^2+cos^2\theta
d\Omega_{n-3}^2\right),
\end{eqnarray}
where $a$ is a parameter associated with its angular momentum, and
$d\Omega^2_{n-3}$ denotes the metric of an unit $(n-3)$ sphere.
The functions $U(r)$, $R(r)$ and $f(r)$ should be determined. In
the particular case $a=0$, this metric reduces to the static and
spherically symmetric cases. For small $a$, we can expect to have
solutions with $U(r)$ still a function of $r$ alone. The $t$
component of the Maxwell equations can be integrated immediately
to give
\begin{equation}\label{Ftr}
F_{tr}=\frac{q e^{4\alpha \Phi /(n-1)}}{\left( rR\right) ^{n-1}} ,
\end{equation}
where $q$ is  an integration constant related to the electric
charge of the solutions. Defining the electric charge via $ Q =
\frac{1}{4\pi} \int \exp\left[{-4\alpha\Phi/(n-1)}\right]  \text{
}^{*} F d{\Omega}, $ we get
\begin{equation}
{Q}=\frac{q\omega _{n-1}}{4\pi},  \label{Charge}
\end{equation}
where $\omega_{n-1}$ represents the area of the unit
$(n-1)$-sphere. In general, in the presence of rotation, there is
also a vector potential in the form
\begin{equation}\label{Aphi}
 A_{\phi}=a q h(r)\sin^2\theta.
\end{equation}
Notice that for infinitesimal rotation parameter, the electric
field (\ref{Ftr}) does not change from the static case. In order
to solve the system of equations (\ref{FE1})-(\ref{FE3}) for four
unknown functions $f(r)$, $R(r)$, $\Phi (r)$ and $h(r)$, we take
the following ansatz
\begin{equation}\label{ansatz}
R(r)=e^{2\alpha \Phi /(n-1)}.
\end{equation}
Inserting (\ref{ansatz}), the Maxwell fields (\ref{Ftr}) and
(\ref{Aphi}), and the metric (\ref{metric}), into the field
equations (\ref{FE1})-(\ref{FE3}), one can show that these
equations have the following solutions
\begin{eqnarray}\label{U}
U(r)&=& -{\frac {(n-2)\left( { \alpha}^{2}+1 \right)
^{2}{b}^{-2\gamma}{r}^{2\gamma}}{\left( { \alpha}^{2}-1 \right)
\left({\alpha}^{2}+n-2 \right) }}-\frac{m}{r^{(n-1)(1-\gamma
)-1}}+\frac{2q^{2}(\alpha ^{2}+1)^{2}b^{-2(n-2)\gamma
}}{(n-1)(\alpha ^{2}+n-2)}r^{2(n-2)(\gamma -1)},
\end{eqnarray}
\begin{eqnarray}\label{f}
f(r)&=&
\frac{m\left({\alpha}^{2}+n-2\right){b}^{(n-3)\gamma}}{{\alpha}^{2}+1}{r}^
{(n-1)(\gamma-1)+1}-\frac{2{q}^{2}( {\alpha}^{2
}+1)b^{(1-n)\gamma}}{n-1}{r}^{2\, \left( n-2 \right)
\left(\gamma-1\right) },
\end{eqnarray}
\begin{eqnarray}\label{Rphi}
\Phi (r)=\frac{(n-1)\alpha }{2(\alpha ^{2}+1)}\ln (\frac{b}{r}),
\hspace{1.5cm} h(r)= r^{(n-3)(\gamma-1)-1}. \label{h}
\end{eqnarray}
Here $b$ is an arbitrary constant and $\gamma =\alpha ^{2}/(\alpha
^{2}+1)$. In the above expressions, $m$ appears as an integration
constant and is related to the ADM (Arnowitt-Deser-Misner) mass of
the black hole. According to the definition of mass due to Abbott
and Deser \cite{abot}, the mass of the solution is \cite{Sheykhi}
\begin{equation}
{M}=\frac{b^{(n-1)\gamma}(n-1) \omega _{n-1}}{16\pi(\alpha^2+1)}m.
\label{Mass}
\end{equation}

\begin{figure}[tbp]
\epsfxsize=7cm \centerline{\epsffile{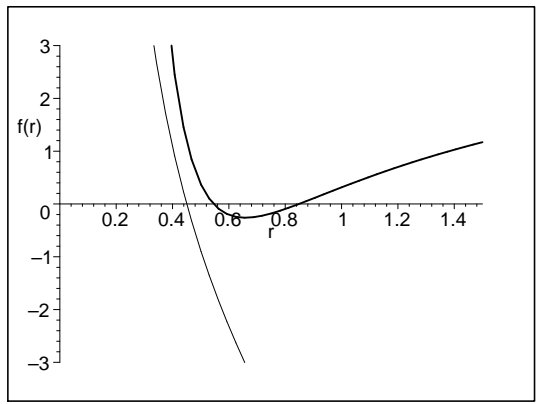}} \caption{The
function $f(r)$ versus $r$ for $q=1$, $b=1$, $m=2$ and $n=4$.
$\protect\alpha=0.5$ (bold line), $\protect\alpha=1.3$ (continuous
line).} \label{figure1}
\end{figure}

\begin{figure}[tbp]
\epsfxsize=7cm \centerline{\epsffile{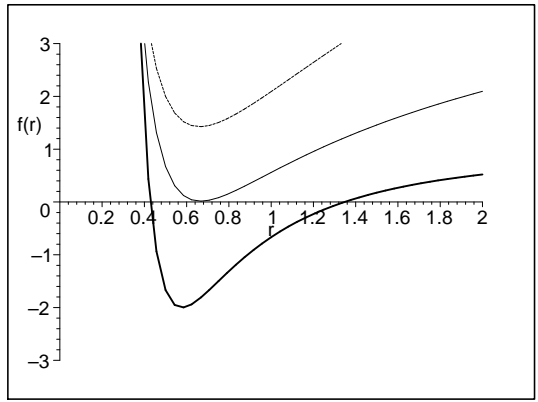}} \caption{The
function $f(r)$ versus $r$ for $q=1$, $b=1$, $m=2$ and $n=4$.
$\protect\alpha=0$ (bold line), $\protect\alpha=0.5$ (continuous
line) and $\protect\alpha=0.7$ (dashed line).} \label{figure2}
\end{figure}

\begin{figure}[tbp]
\epsfxsize=7cm \centerline{\epsffile{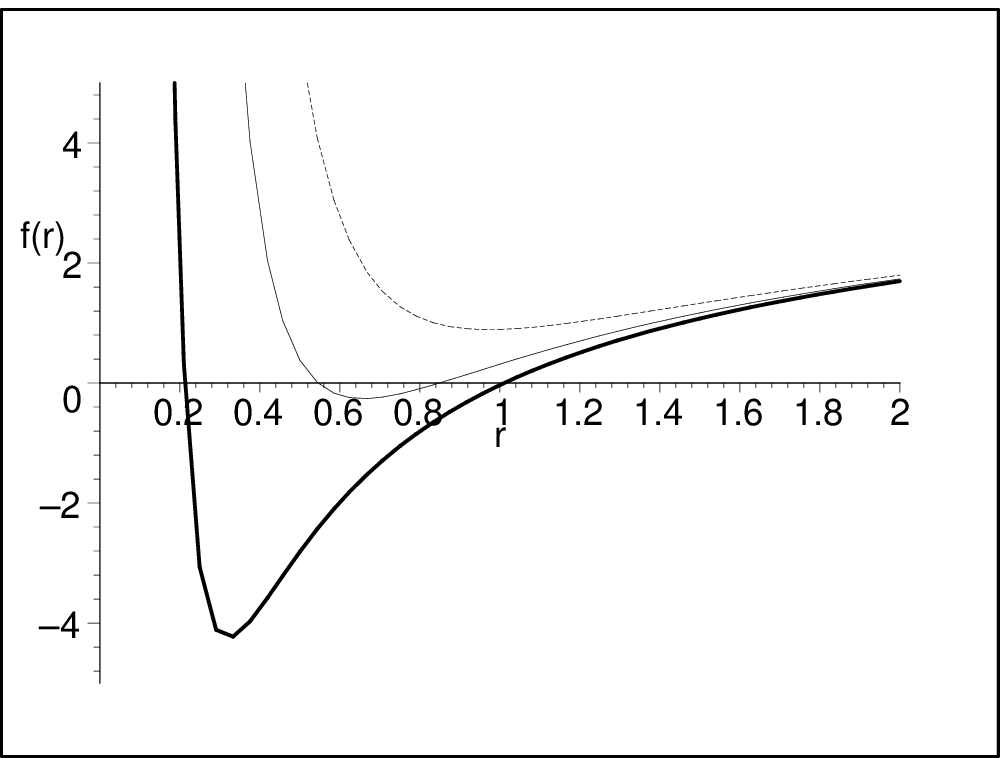}} \caption{The
function $f(r)$ versus $r$ for $\protect\alpha=0.5$, $b=1$, $m=2$
and $n=4$. $q=0.5$ (bold line), $q=1$ (continuous line) and
$q=1.5$ (dashed line).} \label{figure3}
\end{figure}

\begin{figure}[tbp]
\epsfxsize=7cm \centerline{\epsffile{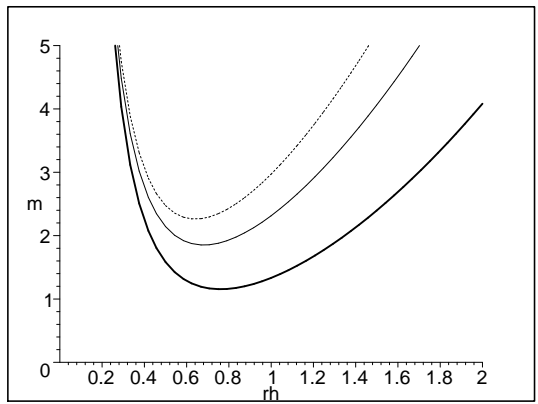}} \caption{The
function $m(r_h)$ versus $r_h$ for $q=1$, $b=1$ and $n=4$.
$\protect\alpha=0$ (bold line), $\protect\alpha=0.5$ (continuous
line) and $\protect\alpha=0.6$ (dashed line).} \label{figure4}
\end{figure}

\begin{figure}[tbp]
\epsfxsize=7cm \centerline{\epsffile{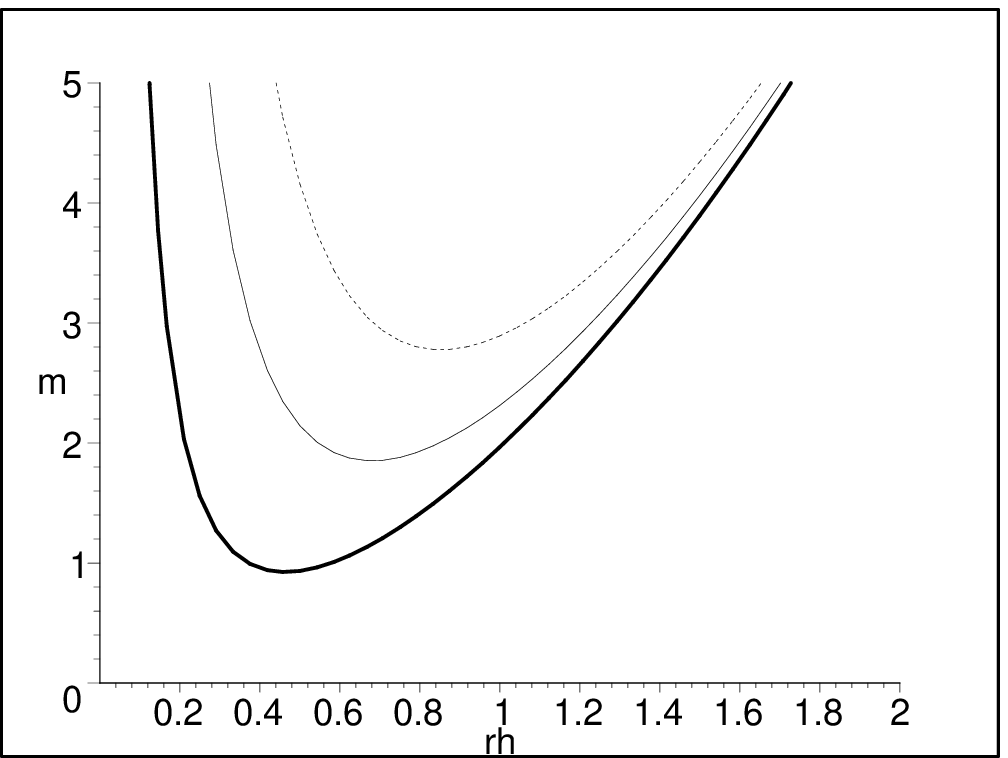}} \caption{The
function $m(r_h)$ versus $r_h$ for $\protect\alpha=0.5$, $b=1$ and
$n=4$. $q=0.5$ (bold line), $q=1$ (continuous line) and $q=1.5$
(dashed line).} \label{figure5}
\end{figure}

For ($\alpha=0=\gamma $) the mass of the black hole reduces to
\begin{equation}\label{m2}
{M}=\frac{(n-1)\omega _{n-1}}{16\pi}m.
\end{equation}
In order to fully satisfy the system of equations, we must have
\begin{equation}\label{lam}
\zeta =\frac{2}{\alpha(n-1)}, \hspace{.8cm} \Lambda
=\frac{(n-1)(n-2)\alpha^2 }{2b^2(\alpha^2-1)}.
\end{equation}
One may also note that in the absence of a non-trivial dilaton
($\alpha=\gamma =0 $), the above solutions reduce to
\begin{eqnarray}
U(r) &=&1-\frac{m}{r^{n-2}}+\frac{2q^2}{(n-1)(n-2)r^{2(n-2)}},
\end{eqnarray}
\begin{eqnarray}\label{f2}
f(r)&=& \frac{m(n-2)}{r^{2-n}}-\frac{2{q}^{2}}{(n-1)r^{2(n-2)}},
\end{eqnarray}
\begin{eqnarray}
h(r)= r^{2-n}, \label{h2}
\end{eqnarray}
which describe an $(n+1)$-dimensional Kerr-Newman black holes in
the limit of slow rotation \cite{Aliev2}. The metric corresponding
to (\ref{U})-(\ref{Rphi}) is neither asymptotically flat nor
(anti)-de Sitter. In order to study the physical properties of
these solutions, we first look for the curvature singularities. In
the presence of a dilaton field, the Kretschmann scalar $R_{\mu
\nu \lambda \kappa }R^{\mu \nu \lambda \kappa }$ diverges at
$r=0$, it is finite for $r\neq 0$ and goes to zero as
$r\rightarrow \infty $. Thus, there is an essential singularity
located at $r=0$. As one can see from (\ref{U}), the solution is
ill-defined for for the string case where $\alpha=1$. The cases
with $\alpha >1$ and $\alpha <1$ should be considered separately.
For $ \alpha >1$, we have  cosmological horizons, while there is
no cosmological horizons if $\alpha <1$ (see fig. \ref{figure1}).
In fact, in the latter case ($\alpha <1$) the spacetimes exhibit a
variety of possible casual structures depending on the values of
the metric parameters (see figs. \ref{figure2}-\ref{figure3}). One
can obtain the casual structure by finding the roots of $ f(r)=0$.
Unfortunately, because of the nature of the exponent in (\ref{U}),
it is not possible to find analytically the location of the
horizons. To get better understanding on the nature of the
horizons, we plot in figures \ref{figure4} and \ref{figure5}, the
mass parameter $m$ as a function of the horizon radius for
different values of dilaton coupling constant $\alpha$. It is easy
to show that the mass parameter $m$ of the dilaton black hole can
be expressed in terms of the horizon radius $r_{h}$ as
\begin{eqnarray}\label{mass}
m(r_{h}) &=&-{\frac {(n-2)( \alpha^{2}+1)
^{2}{b}^{-2\gamma}}{\left( { \alpha}^{2}-1 \right)
\left(n+{\alpha}^{2}-2 \right)
}}{r_{h}}^{n-2+\gamma(3-n)}+\frac{2q^2 \left( {\alpha}^{2}+1
\right) ^{2}{b}^{-2 \gamma(n-2)}}{(n-1) \left(n+{\alpha}^{2}-2
\right)}r_{h}^{(n-3)(\gamma-1)-1}.
\end{eqnarray}

These figures show that for a given value of $\alpha$ or $q$, the
number of horizons depend on the choice of the value of the mass
parameter $m$. We see that, up to a certain value of the mass
parameter $m$, there are two horizons, and as we decrease the $m$
further, the two horizons meet. In this case we get an extremal
black hole with mass $m_{\mathrm{ext}}$. The entropy of the black
hole typically satisfies the so called area law of the entropy
which states that the entropy of the black hole is a quarter of
the event horizon area \cite{Beck}. This near universal law
applies to almost all kinds of black holes, including dilaton
black holes, in Einstein gravity \cite{hunt}. Since the surface
gravity and the area of the event horizon do not change up to the
linear order of the rotating parameter $a$, we can easily show
that the Hawking temperature and the entropy of dilaton black hole
on the outer event horizon $r_{+}$ can be written as
\begin{equation}
T_{+}=\frac{\kappa}{2\pi}= \frac{f^{\text{ }^{\prime
}}(r_{+})}{4\pi},  \hspace{.8cm} S=\frac{\mathcal{A}}{4},
\end{equation}
where $\kappa$ is the surface gravity and $\mathcal{A}$ is the
horizon area. Then, one can get
\begin{eqnarray}\label{Tem}
T_{+}&=&-{\frac {\left( {\alpha}^{2}+n-2 \right) m }{ 4 \pi\left(
{\alpha}^{2}+1 \right) }}{r_{+}}^{ \left( n-1
 \right)  \left( \gamma-1 \right) }+{\frac
{{b}^{-2\,\gamma}\left( n-2 \right) \left( {\alpha}^{2}+1
 \right)   }{2\pi\, \left( 1-{\alpha}^{2} \right)
 }}{r_{+}}^{2\,\gamma-1},
\end{eqnarray}
\begin{equation}
{S}=\frac{b^{(n-1)\gamma}r_{+}^{(n-1)(1-\gamma )}\omega
_{n-1}}{4}.\label{Entropy}
\end{equation}
Equation (\ref{Tem}) shows that for $\alpha >1$ the temperature is
negative. As we argued above in this case we encounter
cosmological horizons, and therefore the cosmological horizons
have negative temperature. Numerical calculations show that the
temperature of the event horizon goes to zero as the black hole
approaches the extreme case. It is a matter of calculation to show
that the mass parameter of the extremal black hole can be written
($\alpha <1$)
\begin{eqnarray}\label{mext}
m_{\mathrm{ext}}&=&{\frac {2\left( n-2 \right)\left(
{\alpha}^{2}+1
 \right) ^{2} {b}^{-2\,\gamma} }{\left( {
1-\alpha}^{2} \right)  \left( {\alpha}^{2}+n-2 \right)
}}{r_{+}}^{(2-n)(\gamma-1)+\gamma},
\end{eqnarray}
In summary, the metric of Eqs. (\ref{metric}) and (\ref{U})-
(\ref{Rphi}) can represent a black hole with inner and outer event
horizons located at $r_{-}$ and $r_{+}$, provided
$m>m_{\mathrm{ext}}$, an extreme black hole in the case of
$m=m_{\mathrm{ext}}$, and a naked singularity if
$m<m_{\mathrm{ext}}$. It is worth noting that in the absence of a
non-trivial dilaton field ($\alpha=\gamma =0 $), expressions
(\ref{Tem}) and (\ref{mext}) reduce to that of an
$(n+1)$-dimensional asymptotically flat black holes.

\begin{figure}[tbp]
\epsfxsize=7cm \centerline{\epsffile{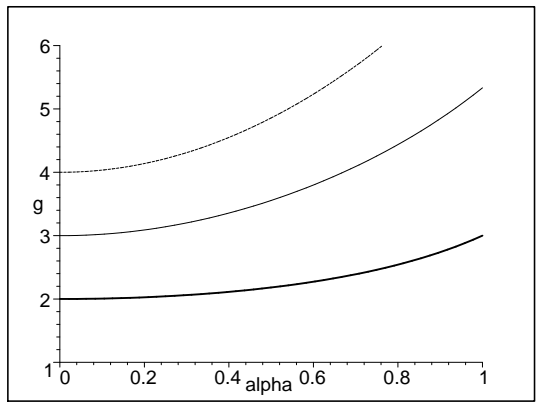}} \caption{The
behaviour of the gyromagnetic ratio $g$ versus $\protect\alpha$
for $b=1$. $n=3$ (bold line), $n=4$ (continuous line) and $n=5$
(dashed line).} \label{figure6}
\end{figure}

\begin{figure}[tbp]
\epsfxsize=7cm \centerline{\epsffile{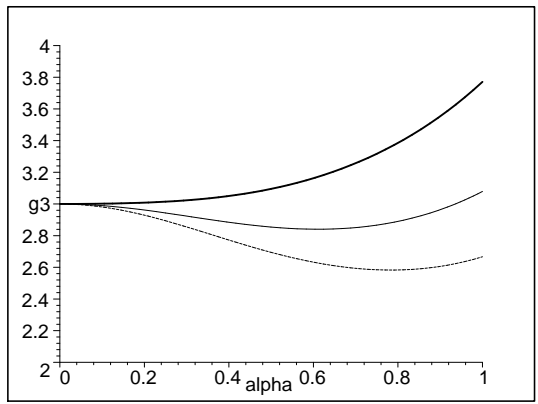}} \caption{The
behaviour of the gyromagnetic ratio $g$ versus $\protect\alpha$
for $n=4$. $b=2$ (bold line), $b=3$ (continuous line) and $b=4$
(dashed line).} \label{figure7}
\end{figure}

Finally, we calculate the angular momentum and the gyromagnetic
ratio of these rotating dilaton black holes which appear in the
limit of slow rotation parameter. The angular momentum of the
dilaton black hole can be calculated through the use of the
quasi-local formalism of the Brown and York \cite{BY}. According
to the quasilocal formalism, the quantities can be constructed
from the information that exists on the boundary of a gravitating
system alone. Such quasilocal quantities will represent
information about the spacetime contained within the system
boundary, just like the Gauss's law. In our case the finite
stress-energy tensor can be written as
\begin{equation}
T^{ab}=\frac{1}{8\pi }\left(\Theta^{ab}-\Theta\gamma ^{ab}\right)
, \label{Stres}
\end{equation}
which is obtained by variation of the action (\ref{act1}) with
respect to the boundary metric $\gamma _{ab}$. To compute the
angular momentum of the spacetime, one should choose a spacelike surface $%
\mathcal{B}$ in $\partial \mathcal{M}$ with metric $\sigma _{ij}$,
and write the boundary metric in ADM form
\[
\gamma _{ab}dx^{a}dx^{a}=-N^{2}dt^{2}+\sigma _{ij}\left( d\varphi
^{i}+V^{i}dt\right) \left( d\varphi ^{j}+V^{j}dt\right) ,
\]
where the coordinates $\varphi ^{i}$ are the angular variables
parameterizing the hypersurface of constant $r$ around the origin,
and $N$ and $V^{i}$ are the lapse and shift functions
respectively. When there is a Killing vector field $\mathcal{\xi
}$ on the boundary, then the quasilocal conserved quantities
associated with the stress tensors of Eq. (\ref{Stres}) can be
written as
\begin{equation}
Q(\mathcal{\xi )}=\int_{\mathcal{B}}d^{n-1}\varphi \sqrt{\sigma }T_{ab}n^{a}%
\mathcal{\xi }^{b},  \label{charge}
\end{equation}
where $\sigma $ is the determinant of the metric $\sigma _{ij}$, $\mathcal{%
\xi }$ and $n^{a}$ are the Killing vector field and the unit
normal vector on the boundary $\mathcal{B}$. For boundaries with
rotational ($\varsigma =\partial /\partial \varphi $) Killing
vector field, one obtains the quasilocal angular momentum
\begin{eqnarray}
J &=&\int_{\mathcal{B}}d^{n-1}\varphi \sqrt{\sigma
}T_{ab}n^{a}\varsigma ^{b},  \label{Angtot}
\end{eqnarray}
provided the surface $\mathcal{B}$ contains the orbits of
$\varsigma $. Finally, the angular momentum of the black holes can
be calculated through the use of Eq. (\ref{Angtot}). We find
\begin{equation}
J
=\frac{(n-\alpha^2)(\alpha^2+n-2)b^{2(n-2)\gamma}\omega_{n-1}}{8\pi
n(n-2) (\alpha^2+1)^2} m a.  \label{J}
\end{equation}
For $a=0$, the angular momentum  vanishes, and therefore $a$ is
the rotational parameter of the dilaton black hole. For
$(\alpha=\gamma=0)$, the angular momentum reduces to the angular
momentum of the $(n+1)$-dimensional Kerr black holes
\begin{equation}
J =\frac{m a \omega_{n-1}}{8\pi} .  \label{J2}
\end{equation}
Combining Eq. (\ref{m2}) with Eq. (\ref{J2}) we get
\begin{equation}
J =\frac{2 M a }{n-1} .  \label{JM}
\end{equation}
Next we calculate the gyromagnetic ratio of this rotating dilaton
black holes. The magnetic dipole moment for this slowly rotating
dilaton black hole is
\begin{equation}
\mu =Q a.
\end{equation}
Therefore, the gyromagnetic ratio is given by
\begin{equation}\label{g}
g=\frac{2\mu M}{QJ} =\frac{n(n-1)(n-2)(\alpha^2+1)}{
(n-\alpha^2)(\alpha^2+n-2)b^{(n-3)\gamma}}.
\end{equation}
It was argued in \cite{Hor1,Ghosh} that the dilaton field can
modify the gyromagnetic ratio of the asymtotically flat and
asymtotically (A)dS four dimensional black holes. Our result here
confirm their arguments. However, in contrast to the gyromagnetic
ratio of the asymtotically flat or (A)dS four dimensional
dilatonic black holes which is turned out to be $g\leq2$, in our
case in which the solutions are neither asymtotically flat nor
(A)dS, we get $g\geq2$ in four dimension. We have shown the
behaviour of the gyromagnetic ratio $g$ of the dilatonic black
hole $(\alpha<1)$ versus $\protect\alpha$ in figures \ref{figure6}
and \ref{figure7}. From these figures we find out that for small
$b$ the gyromagnetic ratio increases with increasing $\alpha$,
while for large $b$ and $n\geq4$ the gyromagnetic ratio decreases
with increasing $\alpha$. In the absence of a non-trivial dilaton
$(\alpha=\gamma=0)$, the gyromagnetic ratio reduces to
\begin{equation}
g=n-1,  \label{J}
\end{equation}
which is the gyromagnetic ratio of the $(n+1)$-dimensional
Kerr-Newman black holes (see e.g. \cite{Aliev2}).

%%%%%%%%%%%%%%%%%%%%%%%%%%%%%%%%%%%%%%%%%%%%%%%%%%%%%%%%%%
\section{Conclusion}
Though the nonrotating black hole solution to the
higher-dimensional Einstein-Maxwell gravity was found several
decades ago \cite{Tan}, the counterpart of the Kerr-Newman
solution in higher dimensions, that is the charged generalization
of the Myers-Perry solution \cite{Myer} in $(n+1)$-dimensional
Einstein-Maxwell gravity, still remains to be found analytically.
Recently, charged rotating black hole solutions in
$(n+1)$-dimensional Einstein-Maxwell theory has been constructed
with a single rotation parameter in the limit of slow rotation
parameter \cite{Aliev2}. In this letter, as a next step to shed
some light on this issue for further investigation, we further
generalized these slowly rotating black hole solutions
(\cite{Aliev2}) by including a dilaton field and a Liouville-type
potential for the dilaton field in the action. These solutions
describe charged rotating dilaton black holes with an arbitrary
dilaton coupling constant in the limit of slow rotation parameter.
In contrast to the rotating black holes in the Einstein-Maxwell
theory, which are asymptotically flat, the charged rotating
dilaton black holes we found here, are neither asymptotically flat
nor (A)dS. Our strategy for constructing these solutions was the
perturbative technique suggested in \cite{Hor1}. We first studied
charged black hole solutions in $(n+1)$-dimensional
Einstein-Maxwell-dilaton gravity. Then, we considered the effect
of adding a small amount of rotation parameter $a$ to the black
hole. We discarded any terms involving $a^2$ or higher power in
$a$. Inspection of the Kerr-Newman solutions shows that the only
term in the metric changes to $O(a)$ is $g_{t\phi}$. Similarly,
the dilaton does not change to $O(a)$ and $A_{\phi}$ is the only
component of the vector potential that change to $O(a)$.  We
showed that in the absence of dilaton field $(\alpha=0=\gamma)$,
our solutions reduce to the $(n+1)$-dimensional Kerr-Newman
modification thereof for small rotation parameter \cite{Aliev2}.
We computed temperature and entropy of black holes, which did not
change to $O(a)$ from the static case. We also obtained the
angular momentum and the gyromagnetic ratio of these rotating
dilaton black holes. Interestingly enough, we found that the
dilaton field can modify the gyromagnetic ratio of the
$(n+1)$-dimensional rotating dilaton black holes. This is in
agrement with the arguments in \cite{Hor1,Ghosh}.

%%%%%%%%%%%%%%%%%%%%%%%%%%%%%%%%%%%%%%%%%%%%%%%%%%%%%%%%%%%%%%%
\acknowledgments{This work has been supported financially by
Research Institute for Astronomy and Astrophysics of Maragha,
Iran.}

%%%%%%%%%%%%%%%%%%%%%%%%%%%%%%%%%%%%%%%%%%%%%%%%%%%%%%%%%%%%%%%%%%%


\begin{thebibliography}{99}


\bibitem{Aliev2} A. N. Aliev, Phys. Rev. D {\bf 74},
024011 (2006).
\bibitem{Strom} A. Strominger and C. Vafa, Phys. Lett. B {\bf 379}, 99
(1996).
 \bibitem{Strom1} J. C. Breckenridge, R. C. Myers, A. W. Peet and C. Vafa,
Phys. Lett. B {\bf 391}, 93 (1997).


\bibitem{RS}  L. Randall, R. Sundrum, Phys. Rev. Lett. {\bf 83}, 3370
(1999); {\em ibid.} {\bf 83}, 4690 (1999).

\bibitem{DGP}  G. Dvali, G. Gabadadze, M. Porrati, Phys. Lett. B {\bf 485},
208 (2000); G. Dvali, G. Gabadadze, Phys. Rev. D {\bf 63} 065007
(2001).

\bibitem{Dim} S. Dimopoulos and G. Landsberg,
Phys. Rev. Lett. {\bf87}, 161602 (2001).
\bibitem{Tan} F. Tangherlini, Nuovo Cimento {\bf27}, 636 (1963).

\bibitem{Myer} R. C. Myers and M. J. Perry, Ann. Phys. (N.Y.) {\bf 172}, 304
(1986).
\bibitem{Emp} R. Emparan and H. S. Reall, Phys. Rev. Lett. {\bf88}, 101101
(2002).

\bibitem{Gau} P. J. Gauntlett, R. C. Myers, and P. K. Townsend,
Classical Quantum Gravity {\bf16}, 1 (1999).

\bibitem{Yo}  D. Youm, Phys. Rep. {\bf316}, 1 (1999);
A.W. Peet, hep-th/0008241.

\bibitem{Cvetic0} M.~Cvetic and D.~Youm, Phys. Rev. D
{\bf54}, 2612 (1996), hep-th/9603147.

\bibitem{Cvetic1} M.~Cvetic and D.~Youm,
Nucl.\ Phys.\  B {\bf 477}, 449 (1996),  hep-th/9605051.

\bibitem{Cvetic2} M.~Cvetic and D.~Youm, Nucl. Phys. B {\bf 476}, 118 (1996)
hep-th/9603100.

\bibitem{Cvetic3} Z. W. Chong, M. Cvetic, H. Lu, and C. N. Pope, Phys. Rev. D
{\bf72}, 041901 (2005); Phys. Rev. Lett. {\bf95}, 161301 (2005).

\bibitem{Aliev3} A. N. Aliev, Mod. Phys. Lett. A {\bf21}, 751 (2006);
A. N. Aliev,  Class. Quant. Gravit. {\bf24}, 4669 (2007).

\bibitem{kunz1}  J. Kunz, F. Navarro-Lérida, A. K. Petersen,
Phys. Lett. B {\bf 614}, 104 (2005).


\bibitem{KimCai} Hyeong-Chan Kim and Rong-Gen Cai, arXiv:0711.0885.

\bibitem{Wit1}  M. B. Green, J. H. Schwarz and E. Witten, {\it Superstring
Theory}, (Cambridge University Press, Cambridge, England, 1987).
\bibitem{Gib} G. W. Gibbons and K. Maeda, Ann. Phys. (N. Y.) {\bf 167},
201 (1986).
\bibitem{Fr}  V. P. Frolov, A. I. Zelnikov and U. Bleyer, Ann. Phys.
(Berlin) {\bf 44}, 371 (1987); Belinsky V and Ruffini R Phys.
Lett. B {\bf 89}, 195 (1980).
\bibitem{Hor1} J. H. Horne and G. T. Horowitz, Phys. Rev. D {\bf 46}, 1340
(1992).
\bibitem{Ghosh} T. Ghosh, S. SenGupta, arXiv: 0709.2754.

\bibitem{Ahar}  O. Aharony, M. Berkooz, D. Kutasov, and N. Seiberg, J. High Energy Phys. \textbf{10},
004 (1998).

\bibitem{DF}  M. H. Dehghani and N. Farhangkhah, Phys. Rev. D {\bf 71},
044008 (2005); M. H Dehghani, Phys. Rev. D {\bf 71}, 064010
(2005).

\bibitem{SDRP}  A. Sheykhi, M. H. Dehghani, N. Riazi and J. Pakravan Phys.
Rev. D {\bf 74}, 084016 (2006).
\bibitem{SDR}  A. Sheykhi, M. H. Dehghani, N. Riazi, Phys.
Rev. D {\bf 75}, 044020 (2007).

\bibitem{DHSR} M. H Dehghani, S. H. Hendi, A. Sheykhi and H. Rastegar Sedehi, JCAP {\bf02} (2007)
020.

\bibitem{CDB1}  G. W. Gibbons and K. Maeda, Nucl. Phys. B {\bf 298}, 741
(1988); T. Koikawa and M. Yoshimura, Phys. Lett. B {\bf 189}, 29
(1987); D. Brill and J. Horowitz, {\em ibid.} B {\bf 262}, 437
(1991).

\bibitem{CDB2}  D. Garfinkle, G. T. Horowitz and A. Strominger, Phys. Rev. D
{\bf 43}, 3140 (1991); R. Gregory and J. A. Harvey, {\em ibid.}
{\bf 47}, 2411 (1993); M. Rakhmanov, {\em ibid.} {\bf 50}, 5155
(1994).
\bibitem{Ast1}  D. Astefanesei, K. Goldstein, S. Mahapatra,
hep-th/0611140; D. Astefanesei, E. Radu Phys. Rev. D {\bf73},
044014, (2006), hep-th/0509144.



\bibitem{MNR} R. B. Mann, E. Radu, C. Stelea, J. High Energy Phys. {\bf0609}, 073
(2006).


\bibitem{MW}  S. Mignemi and D. Wiltshire, Class. Quant. Gravit. {\bf 6},
987 (1989); D. Wiltshire, {Phys. Rev}. D {\bf 44}, 1100 (1991); S.
Mignemi and D. Wiltshire, {\em ibid}. D {\bf 46}, 1475 (1992).

\bibitem{PW}  S. Poletti and D. Wiltshire, Phys. Rev. D {\bf 50}, 7260
(1994); {\em ibid.} {\bf 52}, 3753 (1995); T. Okai, hep-
th/9406126.

\bibitem{CHM}  K. C. K. Chan, J. H. Horne and R. B. Mann, Nucl. Phys. {\bf %
B447}, 441 (1995).


\bibitem{Cai}  R. G. Cai and Y. Z. Zhang,  Phys. Rev. D {\bf 54}, 4891
(1996); R. G. Cai, J. Y. Ji and K. S. Soh, {\em ibid.} {\bf 57},
6547 (1998); R. G. Cai and Y. Z. Zhang, {\em ibid.} {\bf 64},
104015 (2001); R. G.  Cai and A. Wang, Phys. Rev. D {\bf70},
084042 (2004).


\bibitem{Clem}  G. Clement, D. Gal'tsov and C. Leygnac, Phys. Rev. D {\bf67}%
, 024012 (2003); G. Clement and C. Leygnac, {\em ibid}. {\bf 70},
084018 (2004); S. S. Yazadjiev, Class. Quant. Gravit. {\bf22},
3875 (2005); S. S. Yazadjiev, hep-th/0507097; B. Kleihaus, J.
Kunz, F. Navarro-Lerida. Phys. Rev. D {\bf 69}, 064028 (2004).
\bibitem{Sheykhi} A. Sheykhi,  Phys. Rev. D  {\bf76}, 124025 (2007).



\bibitem{kun}  H. Kunduri and J. Lucietti, Phys. Lett. B {\bf 609}, 143
(2005); S. S. Yazadjiev Phys. Rev. D {\bf 72}, 104014 (2005).

\bibitem{kunz2}  J. Kunz, D. Maison, F. N. Lerida and J. Viebahn,
Phys. Lett. B {\bf639} (2006) 95, hep-th/0606005.

\bibitem{Bri} Y. Brihaye, E. Radu, C. Stelea, Class. Quant. Gravit. {\bf24}, 4839
(2007).

\bibitem{Cas}  R. Casadio, B. Harms, Y. Leblanc and P. H. Cox, Phys. Rev. D
{\bf 55}, 814 (1997).

\bibitem{Shi}  K. Shiraishi, Phys. Lett. {\bf A166}, 298 (1992); T. Ghosh
and P. Mitra, Class. Quant. Gravit. {\bf 20}, 1403 (2003).
\bibitem{ShR1}  A. Sheykhi and N. Riazi, Int. J. Mod. Phys. A, Vol. {\bf22}, No. 26, (2007) 4849.
\bibitem{ShR2}  A. Sheykhi and N. Riazi, Int. J. Theor. Phys. {\bf 45},
(2006) 2453.


\bibitem {abot} L. F. Abbott and S. Deser, Nucl. Phys. {\bf B195}, 76 (1982).
\bibitem{Beck}  J. D. Beckenstein, Phys. Rev. D {\bf 7}, 2333 (1973); S. W.
Hawking, Nature (London) {\bf 248}, 30 (1974); G. W. Gibbons and
S. W. Hawking, Phys. Rev. D {\bf 15}, 2738 (1977).
\bibitem{hunt}  C. J. Hunter, Phys. Rev. D {\bf 59}, 024009 (1999); S. W.
Hawking, C. J Hunter and D. N. Page, {\em ibid}. {\bf 59}, 044033
(1999); R. B. Mann {\em ibid}. {\bf 60}, 104047 (1999);{\em ibid}.
{\bf 61}, 084013 (2000).
\bibitem{BY} J. Brown and J. York, Phys. Rev. D {\bf 47}, 1407 (1993).

\end{thebibliography}
\end{document}